\documentclass[prb,twocolumn,amsmath,amsfonts,unsortedaddress,superscriptaddress,floatfix]{revtex4-1}

\usepackage{amssymb}
\usepackage[mathscr]{eucal}
\usepackage[dvips]{graphicx}
\usepackage{color}
\usepackage{bbold}
\usepackage{caption}
\usepackage{subcaption}
\usepackage{hyperref}

\newcommand{\tr}{\textrm{tr}}

\newcommand{\adj}[1]{#1^{\dagger}}

\newcommand{\str}{\textrm{str}}
\newcommand{\eq}[1]{\begin{align}\ensuremath{\displaystyle #1}\end{align}} 
\newcommand{\seq}[1]{\begin{align}\begin{split}\ensuremath{\displaystyle #1}\end{split}\end{align}} 

\newcommand{\ket}[1]{\left\vert{#1}\right\rangle}

\newcommand{\ep}{\epsilon}
\newcommand{\vep}{\varepsilon}
\newcommand{\nb}{\nonumber}
\newcommand{\nes}{^{(1)}}

\renewcommand{\BibitemShut}[1]{}

\begin{document}

\title{\fontsize{12pt}{12pt} \selectfont The nested Algebraic Bethe Ansatz for the supersymmetric t-J and Tensor Networks}

\author{Y. Q. Chong}
\affiliation{C. N. Yang Institute for Theoretical Physics, State University of New York at Stony Brook, NY 11794-3840, USA}

\author{V. Murg}
\affiliation{\mbox{Fakult\"at f\"ur Physik, Universit\"at Wien, Boltzmanngasse 3, A-1090 Vienna, Austria}}

\author{V. E. Korepin}
\affiliation{C. N. Yang Institute for Theoretical Physics, State University of New York at Stony Brook, NY 11794-3840, USA}

\author{F. Verstraete}
\affiliation{\mbox{Fakult\"at f\"ur Physik, Universit\"at Wien, Boltzmanngasse 3, A-1090 Vienna, Austria}}

\date{\today}

\begin{abstract}
We consider a model of strongly correlated electrons in 1D called the t-J model, which was solved by graded algebraic Bethe ansatz.
We use it to design graded tensor networks which can be contracted approximately to obtain a Matrix Product State. As a proof of principle, we calculate observables of ground states and excited states of finite lattices up to $18$ lattice sites.
\end{abstract}

\maketitle

\section{Introduction}

Spin chains have been extensively studied as models for describing quantum systems. For instance, the Heisenberg XXX model was first studied through the means of coordinate Bethe ansatz by Bethe~\cite{bethe31}.

In particular, models of strongly correlated electrons, such as the Hubbard model and t-J model, can also be solved by the Bethe ansatz~\cite{foersterkarowski, esslerkorepin}. In fact, the t-J model is an approximation of the strongly repulsive Hubbard model~\cite{essler05}. These models describe an important physical phenomena: charge and spin separation. The electron becomes unessential, and instead we have spin-waves and holons (holons carry electric charge, but no spin).

The description of quantum states using tensor networks has been very successful in recent literature.
For instance, The extremely successful density matrix normalization group (DMRG)~\cite{white92,white92b} finds its roots in the one-dimensional matrix product states (MPS)~\cite{affleck87,affleck88}.
MPS have also been applied to the field of quantum information
and condensed matter physics~\cite{verstraetecirac05,perezgarcia06,verstraeteciracmurg08,singh10}.
For describing the ground state of higher-dimensional systems,
the projected entangled pair states (PEPS)~\cite{verstraetecirac04} were introduced and
proved to be useful for the numerical study of ground states of two-dimensional systems~\cite{murgverstraete07,murgverstraete08}.
The Multiscale Entanglement Renormalization Ansatz (MERA)~\cite{vidal05,vidal06}
allows the description and numerical study of critical systems.

For the Heisenberg XXX model, it can be easily seen from the tensor network description of the Bethe eigenstates that the eigenstates can be described as MPS:
see also Katsura and Maruyama~[\onlinecite{katsura10}].
Katsura and Maruyama also showed that the alternative formulation of the Bethe Ansatz 
by Alcaraz and Lazo~[\onlinecite{alcaraz04,alcaraz04b,alcaraz06}]
is equivalent to the algebraic Bethe ansatz. Indeed, previous work by three of the co-authors of this paper has managed to use the tensor network formulation of the Heisenberg XXX/XXZ models for periodic and open boundary conditions to obtain correlations for~$50$ sites with good precision~\cite{murg12}.

This paper is devoted to the tensor network description and numerical calculation of observables of the eigenstates of the t-J model.
We first describe the solution of the t-J model, then we proceed with the description of the tensor network and finally we would describe the numerical algorithm used and show the numerical results of the correlation functions.

In order to solve t-J model, the Bethe ansatz (and correspondingly, the tensor network) of the XXX/XXZ model needs to be generalized by two steps: nesting and grading. Nesting means that we first diagonalize the charge degrees of freedom and then the spin degrees of freedom. As such, the Bethe ansatz becomes nested (2 levels) ~\cite{foersterkarowski, esslerkorepin, esmsce}. On the other hand, grading is related to the fermionic nature of the electrons. Graded tensor network states have already been described in the literature~\cite{murgverstraete10}.

Correlation functions are important, but are only described in the double scaling limit for the t-J model~\cite{kawakamiyang90, kawakamiyang91}. The correlation functions have also been described using determinant representations~\cite{zhao06}, but they are highly difficult to evaluate numerically. With the tensor network description of the t-J model, we can investigate correlation functions of eigenstates on finite length lattices for comparison with laboratory results.

\section{Algebraic Bethe ansatz for the t-J model}

In this section, we briefly outline the derivation of the algbraic Bethe ansatz for the t-J model, following Essler and Korepin\cite{esslerkorepin}. 
\subsection{Preliminaries}

In the t-J model, electrons on a lattice of length $L$ are described by operators $c_{j,\sigma}$, $j = 1,\cdots,L$, $\sigma = \pm 1$, which follow the anticommutation relations $\{\adj{c_{i,\sigma}}, c_{j,\tau}\}=\delta_{i,j}\delta_{\sigma,\tau}$. The state $\ket{0}$ (Fock vacuum) satisfies $c_{j,\sigma}\ket{0} = 0$.  The Hilbert space of the Hamiltonian \eqref{hamiltonian1} is constrained to exclude double occupancy, thus there are three possible electronic states at a given lattice site $i$:
\eq{
\ket{0}_i, \ket{\uparrow}_i = \adj{c_{i,1}}\ket{0}_i, \ket{\downarrow}_i = \adj{c_{i,-1}}\ket{0} . \label{allowed_configurations}
}
We define the operators:
\seq{
n_{i,\sigma}&=\adj{c_{j,\sigma}}c_{j,\sigma}, \quad n_i = n_{i,1}+n_{i,-1}, \quad N=\sum_{j=1}^L n_j \\
S_j&=\adj{c_{j,1}} c_{j,-1}, \quad S=\sum_{j=1}^L S_j \\ 
\adj{S_j}&=\adj{c_{j,-1}}c_{j,1}, \quad \adj{S}=\sum_{j=1}^L \adj{S_j} \\ 
S_j^z&=\tfrac12 (n_{j,1} - n_{j,-1}), \quad S^z=\sum_{j=1}^L S_j^z 
}

The t-J Hamiltonian is given by
\eq{
{H} &= \sum_{j=1}^L \left\{ -t \mathcal{P} \sum_{\sigma=\pm 1} (\adj{c_{j,\sigma}}c_{j+1,\sigma} + H.c.) \mathcal{P} \right. \nb \\
&\left. + J(\mathbf{S_j \cdot S_{j+1}} - \tfrac14 n_j n_{j+1}) \right\} \label{hamiltonian1}
}

where $\mathcal{P} = (1-n_{j,-\sigma})$ is the projector which constrains the Hamiltonian to nondoubly occupied states. $t$ represents nearest-neighbor hopping and $J$ represents nearest-neighbor spin exchange and charge interactions.

Adding a term $2N - L$ to the Hamiltonian, and specializing to the value $J=2t=2$, the resultant Hamiltonian is supersymmetric and can be written as a graded permutation operator:
\eq{
{H}_{\textrm{susy}} &= \mathcal{H} + 2N - L \nb \\
&= -\sum_{j=1}^L \Pi^{j,j+1} 
}
The graded permutation operator permutes two adjacent lattice sites as follows (permuting two fermions gives a minus sign):
\begin{align}
\Pi^{j,j+1} \ket{0}_{j} \ket{0}_{j+1} &= \ket{0}_{j} \ket{0}_{j+1} \nb \\
\Pi^{j,j+1} \ket{0}_{j} \ket{\sigma}_{j+1} &= \ket{\sigma}_{j} \ket{0}_{j+1} \\
\Pi^{j,j+1} \ket{\tau}_{j} \ket{\sigma}_{j+1} &= - \ket{\sigma}_{j} \ket{\tau}_{j+1} , \qquad \sigma,\tau = \uparrow, \downarrow \nb 
\end{align}

\subsection{Grading}
Consider the graded linear space $V^{(m\vert n)} = V^m \oplus V^n$, where $m$ and $n$ denote the dimensions of the ``even'' ($V^m$) and ``odd'' ($V^n$) parts, and $\oplus$ denotes the direct sum. Let $\{e_1, \cdots, e_{m+n} \}$ be a basis of $V^{(m+n)}$, such that $\{e_1, \cdots, e_{m} \}$ is a basis of $V^m$ and $\{e_{m+1}, \cdots, e_{n} \}$ is a basis of $V^n$. The Grassmann parities of the basis vectors are given by $\{\ep_1 = \cdots = \ep_{m} = 0 \}$ and $\{\ep_{m+1} = \cdots = \ep_{m+n} = 1 \}$. Linear operators on $V^{(m\vert n)}$ can be represented in block form $[M \in \textrm{End}(V^{(m\vert n)})]:$
\begin{align}
M = \begin{pmatrix}
A & B \\
C & D
\end{pmatrix} , \quad
\ep\begin{pmatrix}
A & 0 \\
0 & D
\end{pmatrix} = 0 , \quad
\ep\begin{pmatrix}
0 & B \\
C & 0
\end{pmatrix} = 1
\end{align}

The supertrace is defined as 
\eq{
\str(M) = \tr(A)-\tr(D),
}
where the traces on the rhs are the usual (non-graded) operator traces in $V^m$ and $V^n$. We now define the graded tensor product of matrices in $V^{(m\vert n)} \otimes V^{(m\vert n)}$  as follows:
\eq{
(F\otimes G)^{ab}_{cd} = F_{ab} G_{cd} (-1)^{\ep_c(\ep_a + \ep_b)}
}
The identity operator $I$ and the permutation operator $\Pi$ are defined as:
\eq{
I_{a_2 b_2}^{a_1 b_1} &= \delta_{a_1 b_1} \delta_{a_2 b_2} \\
\Pi(v \otimes w) &= (w \otimes v), \nb \\
(\Pi)_{a_2 b_2}^{a_1 b_1} &= \delta_{a_1 b_2} \delta_{a_2 b_1} (-1)^{\ep_{b_1} \ep_{b_2}}
}
$V^{(m\vert n)}$ can be interpreted as the space of configurations at every site of a lattice gas of $m$ species of bosons and $n$ species of fermions. For the t-J model, we have $m=1$, $n=2$, and the three allowed configurations are given by \eqref{allowed_configurations}.\\

\subsection{Yang-Baxter equation}
A matrix $R(\lambda)$ fulfills a graded Yang-Baxter equation if the following holds on $V^{(m\vert n)} \otimes V^{(m\vert n)} \otimes V^{(m\vert n)}$:
\eq{
&[I \otimes R(\lambda - \mu)][R(\lambda) \otimes I][I \otimes R(\mu)]\nb \\
&\qquad = [R(\mu) \otimes I][I \otimes R(\lambda)][R(\lambda - \mu) \otimes I] \label{yang_baxter_equation_unbraided}
}
The $R$ matrix
\seq{
&R(\lambda) = b(\lambda) I + a(\lambda) \Pi \\
&a(\lambda) = \frac{\lambda}{\lambda + i}, \quad b(\lambda) = \frac{i}{\lambda + i}
}
is one such matrix that fulfills \eqref{yang_baxter_equation_unbraided}.
We can rewrite \eqref{yang_baxter_equation_unbraided} as
\eq{
&R_{12}(\lambda - \mu) \{[\Pi_{13} R_{13}(\lambda)] \otimes [\Pi_{23} R_{23}(\mu)] \} \nb \\
&\quad = \{[\Pi_{13} R_{13}(\mu)] \otimes [\Pi_{23} R_{23}(\lambda)] \} R_{12}(\lambda - \mu)
\label{yang_baxter_equation_braided}
}
where the indices $1,2,3$ indicate in which of the three tensored spaces the matrices act nontrivially. The tensor product in \eqref{yang_baxter_equation_braided} is between spaces $1$ and $2$. We now call the third space ``quantum space'' and the first two spaces ``matrix spaces''. The quantum space and matrix space are usually called ``physical space'' and ``auxiliary space'' respectively in tensor network terms. The quantum space represents the Hilbert space of a single lattice site. \\
We now define the $L$ operator on site $k$ as a quantum operator valued linear operator on $\mathcal{H}_k \otimes V^{(m\vert n)}_{\textrm{matrix}}$ (where $\mathcal{H}_k \simeq V^{(m\vert n)}$ is the Hilbert space over the $k$th site, and $V^{(m\vert n)}_{\textrm{matrix}}$ is a matrix space):
\eq{
L_k(\lambda)^{ab}_{\alpha \beta} = \Pi^{ac}_{\alpha \gamma} R(\lambda)^{cb}_{\gamma \beta} = [b(\lambda)\Pi + a(\lambda) I]^{ab}_{\alpha \beta} \label{l_operator}.
}
where the Greek (Roman) indices are the ``quantum indices'' (``matrix indices''). Rewriting \eqref{yang_baxter_equation_braided} for the $k$th quantum space,
\eq{
R(\lambda - \mu) [L_k(\lambda)\otimes L_k(\mu)] = [L_k(\mu)\otimes L_k(\lambda)] R(\lambda - \mu)
\label{intertwining_relation_rll_llr}
}
We shall now construct an integrable spin model based on the intertwining relation \eqref{intertwining_relation_rll_llr}. We first define the monodromy matrix $T_L(\lambda)$ as the product (in the matrix space) of the $L$ operators over all of the lattice sites:
\eq{
T_L(\lambda) &= L_L(\lambda) L_{L-1}(\lambda) \cdots L_1(\lambda) \label{monodromy_matrix_lll}
}
$T_L(\lambda)$ is a quantum operator valued $(m+n)\times(m+n)$ matrix that acts nontrivially in the graded tensor product of all quantum spaces of the lattice. It also fulfills the same intertwining relation as the $L$ operators (as can be proven by induction over the length of the lattice):
\eq{
R(\lambda-\mu)[T_L(\lambda) \otimes T_L(\mu)] = [T_L(\mu) \otimes T_L(\lambda)]R(\lambda-\mu) \label{intertwining_relation_rtt_ttr}
}
Taking the supertrace of the monodromy matrix, we get the transfer matrix $\tau(\lambda)$ of the spin model:
\eq{
\tau(\lambda) = \str[T_L(\lambda)] = \sum_{a=1}^{m+n}(-1)^{\ep_a} [T_L(\lambda)]^{aa}
}
As a consequence of \eqref{intertwining_relation_rtt_ttr}, transfer matrices with different spectral parameters commute. This implies that the transfer matrix is the generating functional of the Hamiltonian.
\subsection{Trace identities}
The Hamiltonian \eqref{hamiltonian1} can be obtained from the transfer matrix by taking its first logarithmic derivative at zero spectral parameter and shifting it by a constant:
\seq{
H_{\textrm{susy}} &= -i \left. \frac{\partial \ln[\tau(\lambda)]}{\partial \lambda} \right\vert_{\lambda = 0} -L \\
&= - \sum_{k=1}^{L} (\Pi^{k,k+1}) \label{trace_identity}
}

\subsection{Algebraic Bethe ansatz with FFB grading (Lai representation)}
Let the Hilbert space at the $k$th site of the lattice be spanned by the three vectors $e_1 =(1 0 0)$, $e_2 =(0 1 0)$, and $e_3 =(0 0 1)$. In this section we consider a grading such that $e_1$ and $e_2$ are fermionic and $e_3$ is bosonic, representing the spin-down and spin-up electrons and the empty site respectively. This means that their Grassmann parities are $\ep_1=\ep_2=1$ and $\ep_3=0$. We choose the reference state in the $k$th quantum space $\ket{0}_k$ and the vacuum $\ket{0}$ of the whole lattice to be purely bosonic, i.e.,
\eq{
\ket{0}_n = \begin{pmatrix}
0\\
0\\
1
\end{pmatrix}, \ket{0} = \otimes_{n=1}^L \ket{0}_n
}
This choice of grading implies that $R(\mu) = b(\mu)I + a(\mu)\Pi$ can be written explicitly as:
\begin{widetext}
\eq{
R(\lambda) = \begin{pmatrix}
b(\lambda )-a(\lambda )&0&0&0&0&0&0&0&0\\
0&b(\lambda )&0&-a(\lambda )&0&0&0&0&0\\       
0&0&b(\lambda )&0&0&0&a(\lambda)&0&0\\
0&-a(\lambda )&0&b(\lambda )&0&0&0&0&0\\
0&0&0&0&b(\lambda )-a(\lambda )&0&0&0&0\\
0&0&0&0&0&b(\lambda )&0&a(\lambda )&0\\
0&0&a(\lambda)&0&0&0&b(\lambda )&0&0\\
0&0&0&0&0&a(\lambda )&0&b(\lambda )&0\\
0&0&0&0&0&0&0&0&1
\end{pmatrix} \label{r_matrix_lai}
}
The $L$ operator is defined by \eqref{l_operator} and is of the form
\eq{
L_n(\lambda )=
\left(\begin{array}{ccc}
a(\lambda )-b(\lambda )e_n^{11} &-b(\lambda )e_n^{21} & b(\lambda )e_n^{31}\\
-b(\lambda )e_n^{12}& a(\lambda) -b(\lambda )e_n^{22} & b(\lambda )e_n^{32}\\
b(\lambda )e_n^{13}& b(\lambda )e_n^{23}& a(\lambda )+b(\lambda )e_n^{33}\end{array}
\right), \label{l_operator_lai}}
\end{widetext}
where $(e^{ab}_n)_{\alpha \beta} = \delta_{a\alpha} \delta_{b\beta}$ are quantum operators in the $n$th quantum
space. The monodromy matrix \eqref{monodromy_matrix_lll} can be represented as 
\eq{
T_L(\lambda) &= L_L(\lambda) L_{L-1}(\lambda) \cdots L_1(\lambda) \nb \\
&= 
\left( \begin{array}{ccc}
A_{11}(\lambda )&A_{12}(\lambda ) &B_1(\lambda )\\
A_{21}(\lambda )&A_{22}(\lambda )&B_2(\lambda )\\
C_1(\lambda )&C_2(\lambda )&D(\lambda )\end{array}
\right),
\label{monodromy_matrix_abcd}
}
which is a quantum operator valued $3\times 3$ matrix. For clarity, we write \eqref{monodromy_matrix_abcd} explicitly in component form:
\eq{
\{[T_L(\lambda)]^{ab} \}_{\substack{\alpha_1 \cdots \alpha_L \\ \beta_1 \cdots \beta_L}} = L_L(\lambda)^{ac_L}_{\alpha_L \beta_L} L_{L-1}(\lambda)^{{c_{L}}{c_{L-1}}}_{\alpha_{L-1} \beta_{L-1}} \cdots \nb \\ \cdots 
L_{1}(\lambda)^{c_2c_{1}}_{\alpha_{1} \beta_{1}} (-1)^{\sum_{j=2}^L (\ep_{\alpha_j} + \ep_{\beta_j}) \sum_{i=1}^{j-1} \ep_{\alpha_i}}
\label{monodromy_matrix_component_lai}
}
Note that the physical (greek) indices are subjected to the minus signs from the graded tensor product, while the matrix (latin) indices are not, as they are summed over (and not tensored).
The transfer matrix is then given as
\eq{
\tau(\mu) = \str[T_L(\mu)] = - A_{11}(\mu) - A_{22}(\mu) + D(\mu) \label{transfer_matrix_lai}
}
The action of $L_k(\lambda)$ on $\ket{0}_k$ is
\eq{
L_k(\lambda)\ket{0}_k = \left(\begin{array}{ccc}
a(\lambda)&0&0\\
0&a(\lambda)&0\\
b(\lambda)e_n^{13}&b(\lambda)e_n^{23}&1\end{array}
\right) \ket{0}_k
\label{action_l_operator_lai}
}
Using \eqref{monodromy_matrix_abcd} and \eqref{action_l_operator_lai}, we determine the action of the monodromy matrix on $\ket{0}$ to be
\eq{
T_L(\lambda)\ket{0} = \left(\begin{array}{ccc}
[a(\lambda)]^L&0&0\\
0&[a(\lambda)]^L&0\\
C_1(\lambda)&C_2(\lambda)&1\end{array}
\right) \ket{0}
\label{action_monodromy_matrix_lai}
}
We will now solve for a set of eigenstates of the transfer matrix using the Nested Algebraic Bethe Ansatz (NABA). By inspecting \eqref{action_monodromy_matrix_lai}, $C_1(\lambda)$ and $C_2(\lambda)$ can be interpreted as creation operators (of odd Grassmann parity). We now make the following Ansatz for the eigenstates of $\tau(\mu)$:
\eq{
\ket{\lambda_1, \cdots, \lambda_n \vert F} = C_{a_1}(\lambda_1)C_{a_2}(\lambda_2) \cdots C_{a_n}(\lambda_n) \ket{0} F^{a_n \cdots a_1} , \label{ansatz_state_lai}
}
where $a_j = 1,2$, and $F^{a_n \cdots a_1}$ is a function of the spectral parameters $\lambda$. The action of the transfer matrix on states of the form \eqref{ansatz_state_lai} is determined by \eqref{action_monodromy_matrix_lai} and \eqref{intertwining_relation_rtt_ttr}. The fundamental commutation relations from \eqref{intertwining_relation_rtt_ttr} which are relevant for the NABA are
\seq{
&A_{ab}(\mu)C_c(\lambda) = (-1)^{\ep_a \ep_p} \frac{r(\mu-\lambda)^{dc}_{pb}}{a(\mu - \lambda)} C_p(\lambda)A_{ad}(\mu) \\
&\qquad \qquad \qquad \;\; + \frac{b(\mu-\lambda)}{a(\mu-\lambda)}C_b(\mu)A_{ac}(\lambda),
}
\seq{
&D(\mu)C_c(\lambda) = \frac{1}{a(\lambda-\mu)} C_c(\lambda)D(\mu) - \frac{b(\lambda-\mu)}{a(\lambda-\mu)}C_c(\mu)D(\lambda),\\
&C_{a_1}(\lambda_1)C_{a_2}(\lambda_2) = r(\lambda_1 - \lambda_2)^{b_1 a_2}_{b_2 a_1} C_{b_2}(\lambda_2) C_{b_1}(\lambda_1) , \label{fcr_lai}
}
where
\eq{
r(\mu)^{ab}_{cd}&=b(\mu)\delta_{ab}\delta_{cd} - a(\mu)\delta_{ad}\delta_{bc} \nb\\
&= b(\mu)I^{ab}_{cd} + a(\mu)[\Pi\nes]^{ab}_{cd} \label{r_matrix_nested_lai}
}
Here $[\Pi\nes]^{ab}_{cd} = -\delta_{ad}\delta_{bc}$, is the $4\times 4$ permutation matrix
corresponding to the grading $\ep_1=\ep_2 = 1$. Using \eqref{fcr_lai} we find that the diagonal elements of the monodromy matrix $\tau(\mu)$ act on the states \eqref{ansatz_state_lai} as follows:
\begin{widetext}
\eq{
D(\mu) \ket{\lambda_1, \cdots, \lambda_n \vert F} = \prod_{j=1}^n \frac{1}{a(\lambda_j - \mu)} \ket{\lambda_1, \cdots, \lambda_n \vert F} + \sum_{k=1}^n (\tilde{\Lambda}_k)_{a_1 \cdots a_n}^{b_1 \cdots b_n} C_{b_k}(\mu) \prod_{\substack{j=1\\j\neq k}}^n C_{b_j}(\lambda_j)\ket{0} F^{a_n \cdots a_1} \label{d_on_state_lai},
}
\eq{
[A_{11}(\mu) + A_{22}(\mu)]\ket{\lambda_1, \cdots, \lambda_n \vert F} &= -[a(\mu)]^L \prod_{j=1}^n  \frac{1}{a(\mu - \lambda_j)} \prod_{l=1}^n C_{b_l}(\lambda_l) \ket{0} \tau\nes(\mu)_{a_1 \cdots a_n}^{b_1 \cdots b_n} F^{a_n \cdots a_1} \nb\\ 
&\quad + \sum_{k=1}^n (\Lambda_k)_{a_1 \cdots a_n}^{b_1 \cdots b_n} C_{b_k}(\mu) \prod_{\substack{j=1\\j\neq k}}^n C_{b_j}(\lambda_j)\ket{0} F^{a_n \cdots a_1} \label{a_on_state_lai},
}
\end{widetext}
where we define:
\eq{
L_k\nes&=b(\lambda)\Pi\nes + a(\lambda)I\nes \nb \\
&=\Pi\nes r(\lambda) \nb \\
&= \begin{pmatrix}
a(\lambda)-b(\lambda)e_k^{11} & -b(\lambda) e_k^{21} \\
-b(\lambda) e_k^{12} & a(\lambda)-b(\lambda)e_k^{22}
\end{pmatrix} \label{l_operator_nested_lai} \\
T_n\nes(\mu) &= L_n\nes(\mu-\lambda_n) L_{n-1}\nes(\mu-\lambda_{n-1})  \nb\\
&\quad \cdots L_2\nes(\mu-\lambda_2) L_1\nes(\mu-\lambda_1) \label{monodromy_matrix_nested_lai}\\
&=\begin{pmatrix}
A\nes(\mu) & B\nes(\mu)\\
C\nes(\mu) & D\nes(\mu)\\
\end{pmatrix} , \label{monodromy_matrix_nested_abcd}\\
\tau\nes(\mu)
&= \str[T_n\nes(\mu)]\nb\\
&=- A\nes(\mu) - D\nes(\mu) ,
}
$r(\mu)$ satisfies a (graded) Yang-Baxter equation:
\eq{
r(\lambda-\mu)_{a_3 c_3}^{a_2 c_2} r(\lambda)^{a_1 b_1}_{c_2 d_2} r(\mu)_{c_3 b_3}^{d_2 b_2} = r(\mu)^{a_1 c_1}_{a_2 c_2} r(\lambda)_{a_3 b_3}^{c_2 d_2} r(\lambda-\mu)_{d_2 b_2}^{c_1 b_1}.
\label{yang_baxter_equation_nested_lai}
}
$L\nes$ and $r(\mu)$ can be interpreted as the $L$ operator and $R$ matrix of a fundamental spin model describing two species of fermions. $T_n\nes(\mu)$ and $\tau\nes(\mu)$ are the monodromy matrix and transfer matrix of the corresponding inhomogeneous model. Inspection of \eqref{d_on_state_lai} and \eqref{a_on_state_lai} together with \eqref{transfer_matrix_lai} shows that the eigenvalue condition
\eq{
\tau(\mu) \ket{\lambda_1, \cdots, \lambda_n \vert F} = \nu(\mu,\{\lambda_j\},F)\ket{\lambda_1, \cdots, \lambda_n \vert F} \label{eigenvalue_equation_lai}
}
leads to the requirements that $F$ ought to be an eigenvector of the ``nested'' transfer matrix $\tau\nes(\mu)$, and that the ``unwanted terms'' cancel, i.e.,
\eq{
[-(\Lambda_k)_{a_1 \cdots a_n}^{b_1 \cdots b_n} + (\tilde{\Lambda}_k)_{a_1 \cdots a_n}^{b_1 \cdots b_n}] F^{a_n \cdots a_1} = 0 \label{unwanted_terms_lai}
}
The relative sign in \eqref{unwanted_terms_lai} is due to the supertrace in \eqref{transfer_matrix_lai} and \eqref{eigenvalue_equation_lai}. 
The explicit expressions of $\Lambda_k$ and $\tilde{\Lambda}_k$ can be computed and upon substitution into \eqref{unwanted_terms_lai}, we obtain the following conditions on the spectral parameters $\lambda$, and coefficients $F$, which are necessary for \eqref{eigenvalue_equation_lai} to hold:
\eq{
[a(\lambda_k)]^{-L} &\prod_{\substack{l=1\\l\neq k}}^n \frac{a(\lambda_k - \lambda_l)}{a(\lambda_l - \lambda_k)} F^{b_n \cdots b_1} \nb \\
&= \tau\nes(\lambda_k)^{b_1 \cdots b_n}_{a_1 \cdots a_n} F^{a_n \cdots a_1}, \quad k = 1,\cdots,n \label{bethe_equations0_lai}
}
The first step of the NABA is completed, and we now solve the nesting. The condition that $F$ is an eigenvector of $\tau\nes(\mu)$ requires $\tau\nes(\mu)$ to be diagonalized, which can be achieved by a second, ``nested'' Bethe Ansatz. From \eqref{yang_baxter_equation_nested_lai}, \eqref{l_operator_nested_lai} and \eqref{monodromy_matrix_nested_lai}, the following intertwining relation can be derived:
\eq{
r(\lambda-\mu)[T_L\nes(\lambda) \otimes T_L\nes(\mu)] = [T_L\nes(\mu) \otimes T_L\nes(\lambda)]r(\lambda-\mu) \label{intertwining_relation_nested_rtt_ttr_lai}
}
Using \eqref{monodromy_matrix_nested_abcd}, \eqref{intertwining_relation_nested_rtt_ttr_lai} and \eqref{r_matrix_nested_lai}, we can obtain the nested fundamental commutation relations:
\seq{
D\nes(\mu)C\nes(\lambda) &= \frac{1}{a(\mu - \lambda)} C\nes(\lambda)D\nes(\mu) \\
&\quad - \frac{b(\lambda-\mu)}{a(\lambda-\mu)}C\nes(\mu)D\nes(\lambda),\\
A\nes(\mu)C\nes(\lambda) &= \frac{1}{a(\lambda - \mu)} C\nes(\lambda)A\nes(\mu) \\
&\quad + \frac{b(\mu-\lambda)}{a(\mu-\lambda)}C\nes(\mu)A\nes(\lambda),\\
C\nes(\lambda)C\nes(\mu) &= C\nes(\mu)C\nes(\lambda) . \label{fcr_nested_lai}
}\\
For the nested reference states, we choose:\\
\eq{
\ket{0}_k\nes = \begin{pmatrix}
0\\
1
\end{pmatrix}, \ket{0} = \otimes_{k=1}^n \ket{0}_k\nes
}\\
The action of the nested monodromy matrix $T\nes_n(\mu)$ on the $\ket{0}\nes$ is determined by \eqref{l_operator_nested_lai} and we find 
\seq{
A\nes(\mu)\ket{0}\nes&=\prod_{j=1}^n a(\mu-\lambda_j)\ket{0}\nes \\
D\nes(\mu)\ket{0}\nes&=\prod_{j=1}^n [a(\mu-\lambda_j) - b(\mu-\lambda_j)] \ket{0}\nes \\
&=\prod_{j=1}^n \frac{a(\mu-\lambda_j)}{a(\lambda_j-\mu)}\ket{0}\nes.
}
We now make the following Ansatz for the eigenstates of $\tau\nes(\mu)$
\eq{
\ket{\lambda_1\nes, \cdots, \lambda_{n_1}\nes} = C\nes(\lambda_1\nes)C\nes(\lambda_2\nes) \cdots C\nes(\lambda_{n_1}\nes) \ket{0}\nes , \label{ansatz_state_nested_lai}
}
In component form, this state can be written as $\ket{\lambda_1\nes, \cdots, \lambda_{n_1}\nes}_{a_n \cdots a_1}$, which is directly identifiable with $F^{a_n \cdots a_1}$.\\
The action of $\tau\nes(\mu)$ on the states \eqref{ansatz_state_nested_lai} can be evaluated with the help of the nested fundamental commutation relations \eqref{fcr_nested_lai}:
\begin{widetext}
\eq{
D\nes(\mu) \ket{\lambda_1\nes, \cdots, \lambda_{n_1}\nes} = \prod_{j=1}^{n_1} \frac{1}{a(\mu - \lambda_j\nes)} \prod_{l=1}^{n} \frac{a(\mu - \lambda_l)}{a(\lambda_l - \mu)} \ket{\lambda_1\nes, \cdots, \lambda_{n_1}\nes}
+ \sum_{k=1}^{n_1} \tilde{\Lambda}_k\nes C\nes(\mu) \prod_{\substack{j=1\\j\neq k}}^n C\nes(\lambda_j\nes)\ket{0}\nes \label{d_on_state_nested_lai},\\
A\nes(\mu) \ket{\lambda_1\nes, \cdots, \lambda_{n_1}\nes} = \prod_{j=1}^{n_1} \frac{1}{a(\lambda_j\nes-\mu)} \prod_{l=1}^{n} a(\mu - \lambda_l) \ket{\lambda_1\nes, \cdots, \lambda_{n_1}\nes}
+ \sum_{k=1}^{n_1} {\Lambda}_k\nes C\nes(\mu) \prod_{\substack{j=1\\j\neq k}}^n C\nes(\lambda_j\nes)\ket{0}\nes \label{a_on_state_nested_lai}.
}
From \eqref{a_on_state_nested_lai} and \eqref{d_on_state_nested_lai} one can read off the eigenvalues of $\tau\nes(\mu)$:
\eq{
\tau\nes(\mu) \ket{\lambda_1\nes, \cdots, \lambda_{n_1}\nes} = - \left[ 
\prod_{j=1}^{n_1} \frac{1}{a(\mu - \lambda_j\nes)} \prod_{l=1}^{n} \frac{a(\mu - \lambda_l)}{a(\lambda_l - \mu)} 
+ \prod_{j=1}^{n_1} \frac{1}{a(\lambda_j\nes-\mu)} \prod_{l=1}^{n} a(\mu - \lambda_l) 
\right]\ket{\lambda_1\nes, \cdots, \lambda_{n_1}\nes} \label{eigenvalues_nested_transfer_matrix_lai}.
}
\end{widetext}
Substituting \eqref{eigenvalues_nested_transfer_matrix_lai} into \eqref{bethe_equations0_lai} at $\mu=\lambda_k$, we obtain the first of Bethe equations
\eq{
[a(\lambda_k)]^L = \prod_{i=1}^{n_1} a(\lambda_k - \lambda_i\nes), k = 1,\cdots,n . \label{bethe_equations1_lai}
}
The explicit expressions of $\Lambda_k$ and $\tilde{\Lambda}_k$ can be computed and their cancellation [to ensure that the states \eqref{ansatz_state_nested_lai} are eigenstates of the transfer matrix
$\tau\nes(\mu)$] leads to the following set of Bethe equations for the nesting:
\eq{
\prod_{i=1}^n a(\lambda_i - \lambda_p\nes) = \prod_{\substack{j=1\\j\neq p}}^{n_1} \frac{a(\lambda_j\nes-\lambda_p\nes)}{a(\lambda_p\nes-\lambda_j\nes)}, \,\, p = 1,\cdots,n_1.\label{bethe_equations2_lai}
}
Due to our choice of grading, we find that $n=N_e=N_\uparrow+N_\downarrow$ and $n_1=N_\downarrow$. If we define the shifted spectral parameters $\tilde{\lambda}_k=\lambda_k+i/2$, we can rewrite the Bethe equations in their ``generic'' form:
\seq{
&\left[\frac{\tilde{\lambda}_k - i/2}{\tilde{\lambda}_k + i/2}\right]^L = \prod_{j=1}^{N_\downarrow} \frac{\tilde{\lambda}_k - \lambda_j\nes - i/2}{\tilde{\lambda}_k - \lambda_j\nes + i/2}, \,\, k = 1,\cdots,N_e \\
&\prod_{k=1}^{N_e} \frac{\tilde{\lambda}_k - \lambda_p\nes - i/2}{\tilde{\lambda}_k - \lambda_p\nes + i/2} = \prod_{\substack{j=1\\j\neq p}}^{N_\downarrow} \frac{\lambda_j\nes - \lambda_p\nes - i}{\lambda_j\nes - \lambda_p\nes + i}, \,\, p = 1,\cdots,n_1 \label{bethe_equations3_lai}
}
The eigenvalues of the transfer matrix \eqref{transfer_matrix_lai} are given by
\seq{
&\nu(\mu,\{\lambda_j\},F)=[a(\mu)]^L \prod_{j=1}^{N_e} \frac{1}{a(\mu-\lambda_j)} \nu\nes(\mu) \\
&\qquad \qquad \qquad + \prod_{j=1}^{N_e} \frac{1}{a(\lambda_j-\mu)}\\
&\nu\nes(\mu) = - \left( \prod_{i=1}^{N_\downarrow} \frac{1}{a(\mu - \lambda_i\nes)} \prod_{j=1}^{N_e} \frac{a(\mu-\lambda_j)}{a(\lambda_j-\mu)} \right. \\
&\qquad \qquad \qquad \left. + \prod_{i=1}^{N_h} \frac{1}{a(\lambda_i\nes-\mu)} \prod_{j=1}^{N_e} a(\mu-\lambda_j) \right) .\label{eigenvalues_transfer_matrix_lai}
}
Using the trace identities \eqref{trace_identity}, we can obtain the energy eigenvalues from the eigenvalues of the transfer matrix:
\seq{
E_{\textrm{susy}} &= \sum_{j=1}^{N_e} \frac{1}{\tilde{\lambda}_j^2 + 1/4} - L \\
& = -2 \sum_{j=1}^{N_e} \cos(k_j) + 2N_e - L, \label{energy_lai}
}
where we have reparameterized $\tilde{\lambda}_j = \tfrac12 \cot(k_j/2)$. The Bethe equations \eqref{bethe_equations3_lai} and the energy \eqref{energy_lai} were also derived by Schlottmann\cite{tj_schlottmann} and Lai\cite{tj_lai} independently.

\subsection{Algebraic Bethe ansatz with BFF grading (Sutherland representation)}
In this section we consider a grading such that $e_2$ and $e_3$ are fermionic and $e_1$ is bosonic, representing the spin-down and spin-up electrons and the empty site respectively. This means that their Grassmann parities are $\ep_2=\ep_3=1$ (fermionic) and $\ep_1=0$ (bosonic). We choose the reference state in the $k$th quantum space $\ket{0}_k$ and the vacuum $\ket{0}$ of the whole lattice to be fermionic with all spins up, i.e.,
\eq{
\ket{0}_n = \begin{pmatrix}
0\\
0\\
1
\end{pmatrix}, \ket{0} = \otimes_{n=1}^L \ket{0}_n
}
This choice of grading implies that $R$ can be written as
\begin{widetext}
\eq{
R(\lambda) = \begin{pmatrix}
1&0&0&0&0&0&0&0&0\\
0&b(\lambda )&0&a(\lambda )&0&0&0&0&0\\       
0&0&b(\lambda )&0&0&0&a(\lambda)&0&0\\
0&-a(\lambda )&0&b(\lambda )&0&0&0&0&0\\
0&0&0&0&b(\lambda )a(\lambda )&0&0&0&0\\
0&0&0&0&0&b(\lambda )&0&-a(\lambda )&0\\
0&0&a(\lambda)&0&0&0&b(\lambda )&0&0\\
0&0&0&0&0&-a(\lambda )&0&b(\lambda )&0\\
0&0&0&0&0&0&0&0&b(\lambda )-a(\lambda )
\end{pmatrix} \label{r_matrix_suth}
}
The $L$ operator is
\eq{
L_n(\lambda )=
\left(\begin{array}{ccc}
a(\lambda )+b(\lambda )e_n^{11} &b(\lambda )e_n^{21} & b(\lambda )e_n^{31}\\
b(\lambda )e_n^{12}& a(\lambda) -b(\lambda )e_n^{22} & -b(\lambda )e_n^{32}\\
b(\lambda )e_n^{13}& -b(\lambda )e_n^{23}& a(\lambda )-b(\lambda )e_n^{33}\end{array}
\right), \label{l_operator_suth}}
The action of $L_k(\lambda)$ on $\ket{0}_k$ is
\eq{
L_k(\lambda)\ket{0}_k = \left(\begin{array}{ccc}
a(\lambda)&0&0\\
0&a(\lambda)&0\\
b(\lambda)e_n^{13}&-b(\lambda)e_n^{23}&a(\lambda)-b(\lambda)\end{array}
\right) \ket{0}_k
\label{action_l_operator_suth}
}
The monodromy matrix is partitioned as before in \eqref{monodromy_matrix_abcd}, which now gives the transfer matrix
\eq{
\tau(\mu) = A_{11}(\mu) - A_{22}(\mu) - D(\mu) \label{transfer_matrix_suth}
}
The action of the monodromy matrix on $\ket{0}$ follows from \eqref{action_l_operator_suth}:
\eq{
T_L(\lambda)\ket{0} = \left(\begin{array}{ccc}
[a(\lambda)]^L&0&0\\
0&[a(\lambda)]^L&0\\
C_1(\lambda)&C_2(\lambda)&[a(\lambda)-b(\lambda)]^L\end{array}
\right) \ket{0}
\label{action_monodromy_matrix_suth}
}
and by inspecting \eqref{action_monodromy_matrix_suth}, $C_1(\lambda)$ and $C_2(\lambda)$ are found to be creation operators of odd and even Grasssmann parity respectively.
We make the following Ansatz for the eigenstates of $\tau(\mu)$:
\eq{
\ket{\lambda_1, \cdots, \lambda_n \vert F} = C_{a_1}(\lambda_1)C_{a_2}(\lambda_2) \cdots C_{a_n}(\lambda_n) \ket{0} F^{a_n \cdots a_1} , \label{ansatz_state_suth}
}
The fundamental commutation relations are found to be
\seq{
A_{ab}(\mu)C_c(\lambda) &= (-1)^{\ep_a \ep_p + \ep_a + \ep_b} \frac{r(\mu-\lambda)^{dc}_{pb}}{a(\mu - \lambda)} C_p(\lambda)A_{ad}(\mu) \\
&\quad + (-1)^{(\ep_a + 1)(\ep_b + 1)}  \frac{b(\mu-\lambda)}{a(\mu-\lambda)}C_b(\mu)A_{ac}(\lambda),\\
D(\mu)C_c(\lambda) &= \frac{1}{a(\lambda-\mu)} C_c(\lambda)D(\mu) - \frac{b(\lambda-\mu)}{a(\lambda-\mu)}C_c(\mu)D(\lambda),\\
C_{a_1}(\lambda_1)C_{a_2}(\lambda_2) &= r_{FB}(\lambda_1 - \lambda_2)^{a_2 b_1}_{a_1 b_2} C_{b_2}(\lambda_2) C_{b_1}(\lambda_1) , \label{fcr_suth}
}
where
\seq{
r(\mu)_{cd}^{ab} = b(\mu)I_{cd}^{ab} + a(\mu)(\Pi_{BF})_{cd}^{ab}, \quad
r_{FB}(\mu)_{cd}^{ab} = b(\mu)I_{cd}^{ab} + a(\mu)(\Pi_{FB})_{cd}^{ab}, \label{r_matrix_nested_suth}
}
and $\Pi_{BF}$ and $\Pi_{FB}$ are the permutation matrices for the gradings $\ep_1=0$, $\ep_2=1$ and $\ep_1=1, \ep_2=0$, respectively. Using \eqref{fcr_suth} we find that the diagonal elements of the monodromy matrix act on the states \eqref{ansatz_state_suth} as follows:
\eq{
D(\mu) \ket{\lambda_1, \cdots, \lambda_n \vert F} = \prod_{j=1}^n \frac{1}{a(\lambda_j - \mu)} \left(\frac{a(\mu)}{a(-\mu)}\right)^L \ket{\lambda_1, \cdots, \lambda_n \vert F} + \sum_{k=1}^n (\tilde{\Lambda}_k)_{a_1 \cdots a_n}^{b_1 \cdots b_n} C_{b_k}(\mu) \prod_{\substack{j=1\\j\neq k}}^n C_{b_j}(\lambda_j)\ket{0} F^{a_n \cdots a_1} \label{d_on_state_suth},
}
\eq{
[A_{11}(\mu) - A_{22}(\mu)]\ket{\lambda_1, \cdots, \lambda_n \vert F} &= [a(\mu)]^L \prod_{j=1}^n  \frac{1}{a(\mu - \lambda_j)} \prod_{l=1}^n C_{b_l}(\lambda_l) \ket{0} \tau\nes(\mu)_{a_1 \cdots a_n}^{b_1 \cdots b_n} F^{a_n \cdots a_1} \nb\\ 
&\quad + \sum_{k=1}^n (\Lambda_k)_{a_1 \cdots a_n}^{b_1 \cdots b_n} C_{b_k}(\mu) \prod_{\substack{j=1\\j\neq k}}^n C_{b_j}(\lambda_j)\ket{0} F^{a_n \cdots a_1} \label{a_on_state_suth},
}
where
\eq{
\tau\nes(\mu)_{a_1 \cdots a_n}^{b_1 \cdots b_n} = (-1)^{\ep_c} L_n\nes(\mu-\lambda_n)^{cc_{n-1}}_{b_n a_n} L_{n-1}\nes(\mu-\lambda_{n-1})^{c_{n-1}c_{n-2}}_{b_{n-1} a_{n-1}} \cdots L_1\nes(\mu-\lambda_1)^{c_1c}_{b_1 a_1} (-1)^{\ep_c \sum_{i=1}^{n-1} (\ep_{b_i}+1)\sum_{i=1}^{n-1} \ep_{c_i}(\ep_{b_i}+1)}, \label{monodromy_matrix_nested_component_suth}
}
Here all the indices $c_i$ and $c$ are summed over. $\tau\nes(\mu)$ is the transfer matrix of an inhomogeneous spin model of a boson and fermion on a lattice of $n$ sites. Our reference state $\ket{0}$ is now of fermionic nature and we have to define a graded tensor product reflecting this fact:
\eq{
(F\overline{\otimes} G)^{ab}_{cd} = F_{ab} G_{cd} (-1)^{(\ep_c+1)(\ep_a + \ep_b)}
}
In terms of this tensor product, the transfer matrix $\tau\nes(\mu)$ given by \eqref{intertwining_relation_nested_rtt_ttr_suth} can be obtained as
\eq{
\tau\nes(\mu)_{a_1 \cdots a_n}^{b_1 \cdots b_n} &= \str[T_n\nes(\mu)]\nb\\
&=\str[L_n\nes(\mu-\lambda_n) \overline{\otimes} L_{n-1}\nes(\mu-\lambda_{n-1}) \overline{\otimes} \cdots  \overline{\otimes} L_1\nes(\mu-\lambda_1)] \label{monodromy_matrix_nested_suth}, \\
L_k\nes&=b(\lambda)\Pi\nes_{BF} + a(\lambda)I\nes = \begin{pmatrix}
a(\lambda)+b(\lambda)e_k^{11} & b(\lambda) e_k^{21} \\
b(\lambda) e_k^{12} & a(\lambda)-b(\lambda)e_k^{22}
\end{pmatrix} \label{l_operator_nested_suth}
}
\end{widetext}

In \eqref{monodromy_matrix_nested_suth} we have explicitly written the tensor product $\overline{\otimes}$ between the quantum spaces over the sites of the inhomogeneous model (and the $L$ operators are multiplied within the matrix space). As before, $F^{a_n \cdots a_1}$
must be an eigenvector of $\tau\nes(\mu)$ if $\ket{\lambda_1, \cdots, \lambda_n \vert F}$ is to be
an eigenstate of $\tau(\mu)$. The unwanted terms can be computed in a similar way to the ones described for the FFB grading. The condition of the cancellation of the unwanted terms,
\eq{
[(\Lambda_k)_{a_1 \cdots a_n}^{b_1 \cdots b_n} - (\tilde{\Lambda}_k)_{a_1 \cdots a_n}^{b_1 \cdots b_n}] F^{a_n \cdots a_1} = 0, \label{unwanted_terms_suth}
}
leads to the conditions
\eq{
F^{a_n \cdots a_1} = [a(-\lambda_k)]^L[\tau\nes(\lambda_k)F]^{a_n \cdots a_1}, \,\, k = 1,\cdots,n .
}
To solve the nesting we first have to note that, due to our change of tensor product, the nested $L$ operators $L\nes(\lambda)$ are now interwined by the $R$ matrix
\eq{
\widehat{r}(\mu)_{cd}^{ab} = b(\mu)\delta_{ab}\delta_{cd} + a(\mu)\delta_{ad}\delta_{bc}(-1)^{\ep_a+\ep_c+\ep_a \ep_c}. \label{r_matrix_nested1_suth}
}
The intertwining relation
\eq{
\widehat{r}(\lambda-\mu)[T_L\nes(\lambda) \overline{\otimes} T_L\nes(\mu)] = [T_L\nes(\mu) \overline{\otimes} T_L\nes(\lambda)]\widehat{r}(\lambda-\mu) \label{intertwining_relation_nested_rtt_ttr_suth}
}
together with the choice of vacuum,
\eq{
\ket{0}_k\nes = \begin{pmatrix}
0\\
1
\end{pmatrix}, \ket{0} = \overline{\otimes}_{k=1}^n \ket{0}_k\nes
}
can be analyzed similar to what was done in previous section. It can be shown that they represent a model of the permutation type with $BF$ grading. The resulting Bethe equations are
\eq{
&[a(-\lambda_l)]^L = \prod_{\substack{m=1\\m\neq l}}^{n} \frac{a(\lambda_m - \lambda_l)}{a(\lambda_l - \lambda_m)} \prod_{j=1}^{n_1} a(\lambda_k - \lambda_i\nes),\nb\\& \qquad \qquad \qquad \qquad \qquad \qquad \qquad \qquad l = 1,\cdots,n . \label{bethe_equations1_suth} \\
&1 = \prod_{j=1}^n a(\lambda_j - \lambda_k\nes), \,\, k = 1,\cdots,n_1.\label{bethe_equations2_suth}
}
Due to our choice of grading, we find that $n=N_h+N_\downarrow$ and $n_1=N_h$ respectively, where $N_h=N-N_e$ is the number of holes. If we define the shifted spectral parameters
\eq{
\tilde{\lambda}_j=\lambda_j-i/2, \,\, \tilde{\lambda}_j\nes=\lambda_j\nes-i,
}
we obtain Sutherland's \cite{tj_suth} form of the periodic boundary conditions: 
\seq{
&\left[\frac{\tilde{\lambda}_k - i/2}{\tilde{\lambda}_k + i/2}\right]^L = \prod_{\substack{m=1\\m\neq l}}^{N_h + N_\downarrow} \frac{\tilde{\lambda}_l - \tilde{\lambda}_m - i}{\tilde{\lambda}_l - \tilde{\lambda}_m + i}
\prod_{j=1}^{N_h} \frac{\tilde{\lambda}_l - \tilde{\lambda}_j\nes - i/2}{\tilde{\lambda}_l - \tilde{\lambda}_j\nes + i/2},\\
&\qquad \qquad \qquad \qquad \qquad \qquad \qquad \qquad l = 1,\cdots,{N_h + N_\downarrow},  \\
&1 = \prod_{k=1}^{N_h + N_\downarrow} \frac{\tilde{\lambda}_j - \tilde{\lambda}_k\nes - i/2}{\tilde{\lambda}_j - \tilde{\lambda}_k\nes + i/2}, \,\, k = 1,\cdots,N_h \label{bethe_equations3_suth}
}
The eigenvalues of the transfer matrix are
\seq{
&\nu(\mu,\{\lambda_j\},F)=[a(\mu)]^L \prod_{j=1}^{N_h + N_\downarrow} \frac{1}{a(\mu-\lambda_j)} \nu\nes(\mu) \\
&\qquad \qquad \qquad - \prod_{j=1}^{N_h + N_\downarrow} \frac{1}{a(\mu-\lambda_j)} \left(\frac{a(\mu)}{a(-\mu)}\right)^L\\
&\nu\nes(\mu) = \prod_{l=1}^{N_h} \frac{1}{a(\mu - \lambda_j\nes)} \left(\prod_{j=1}^{N_h + N_\downarrow} a(\mu-\lambda_j) \right. \\
&\qquad \qquad \qquad \qquad \qquad \qquad \left. - \prod_{j=1}^{N_h + N_\downarrow} \frac{a(\mu-\lambda_j)}{a(\lambda_j-\mu)} \right) \label{eigenvalues_transfer_matrix_suth}
}
Using the trace identities \eqref{trace_identity}, we can obtain the energy eigenvalues as:
\seq{
E_{\textrm{susy}} &= L - \sum_{j=1}^{N_h + N_\downarrow} \frac{1}{\tilde{\lambda}_j^2 + 1/4} \\
& = L - 2(N_h+N_\downarrow) - 2 \sum_{j=1}^{N_e} \cos(k_j) , \label{energy_suth}
}
where we have reparameterized $\tilde{\lambda}_j = \tfrac12 \tan(k_j/2)$

\section{Tensor network description of the Bethe ansatz}

\subsection{Tensor network form}
\begin{figure}[h]
\begin{subfigure}[h]{0.1\textwidth}
  \includegraphics[width=\textwidth]{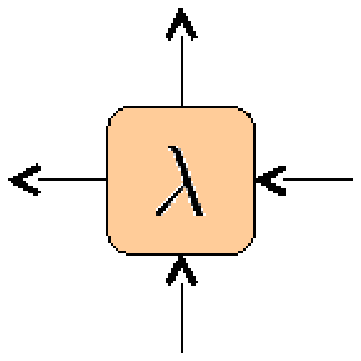} \caption{\label{fig:l_op}$L$ operator $L(\lambda)$} \quad
\end{subfigure}
\begin{subfigure}[h]{0.35\textwidth}
  \includegraphics[width=\textwidth]{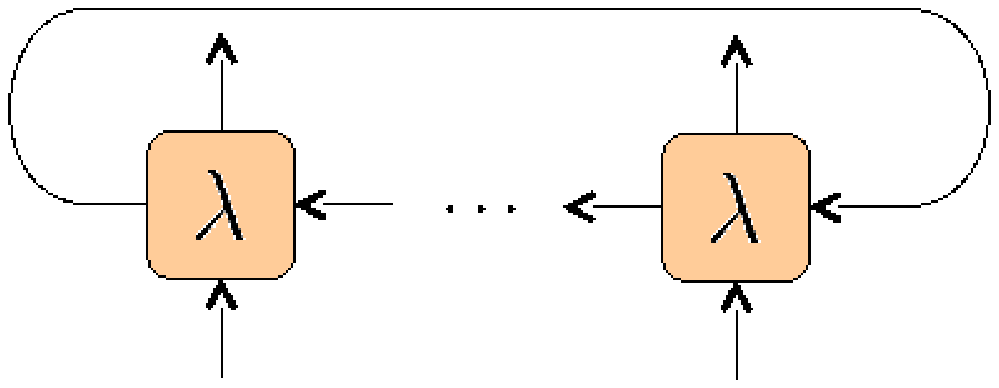} \caption{\label{fig:transfer_matrix}Monodromy matrix $T(\lambda)$}
\end{subfigure}
\end{figure}

\begin{figure}[h]
    \includegraphics[width=0.45\textwidth]{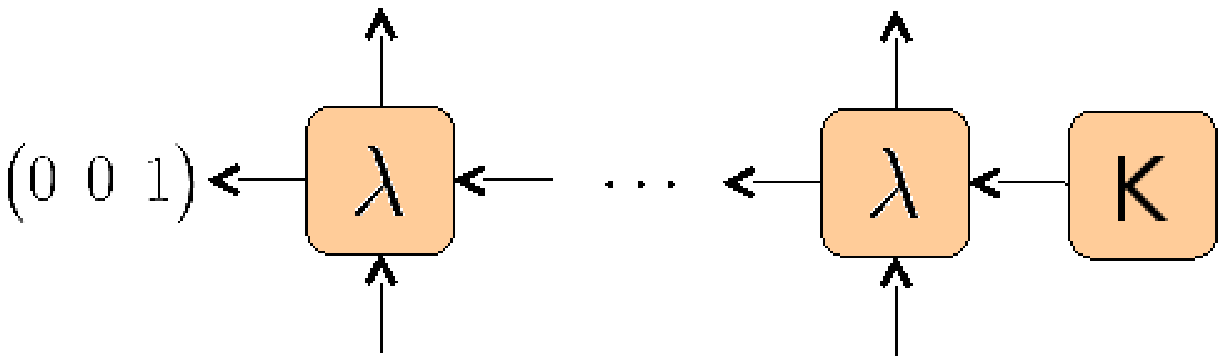}
    \caption{\label{fig:creation_operator}Creation operators $\{C_1(\lambda),C_2(\lambda)\}$}
\end{figure}

\begin{figure}[h]
    \includegraphics[width=0.1\textwidth]{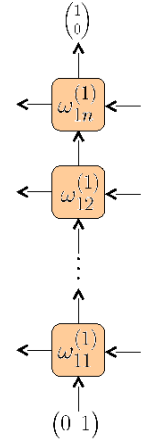}
    \caption{\label{fig:creation_operator_nested}Nested creation operator $C\nes(\lambda)$}
\end{figure}

We now represent the above NABA in tensor network form. If we leave the considerations for grading aside, the (abstract) form of the tensor network is the same for both Lai and Sutherland representation (only actual mathematical representation differs). We proceed below to consider the general form of the tensor network for both representations without considering the grading first, after which we then consider the grading in detail in Sec.~\ref{sec:grading}.\\
We represent each $L$ operator $L(\lambda)_{\alpha\beta}^{ab}$ (a tensor with four indices) as shown in Fig.~\ref{fig:l_op}. We construct the transfer matrix $T_L(\lambda) = L_L(\lambda) L_{L-1}(\lambda) \cdots L_1(\lambda)$ as shown in Fig.~\ref{fig:transfer_matrix}.\\
For the first level Bethe ansatz, the set of creation operators $\{C_1,C_2\}$ in \eqref{monodromy_matrix_abcd} is constructed by terminating the ends of the transfer matrix by boundary vectors/matrices as shown in Fig.~\ref{fig:creation_operator}. The boundary row vector $(0 0 1)$ on the left selects the third row of the transfer matrix $T(\lambda)$. The matrix $K$, which selects the first and second column of $T(\lambda)$, is defined as:
\eq{
K = \begin{pmatrix}
1&0\\
0&1\\
0&0
\end{pmatrix} \label{connector_k}
}
We call the matrix $K$ the connector for it will be the bridge between the first level and nested Bethe ansatz. \\
For the nested Bethe ansatz, the creation operator $C\nes(\lambda)$ in \eqref{monodromy_matrix_nested_abcd} is constructed by terminating the ends of the transfer matrix by boundary vectors $(0\;1)$ on the left and $(1\;0)^\intercal$ on the right (selecting the second row and first column respectively) as shown in Fig.~\ref{fig:creation_operator_nested}.

\begin{figure}[h]
    \includegraphics[width=0.45\textwidth]{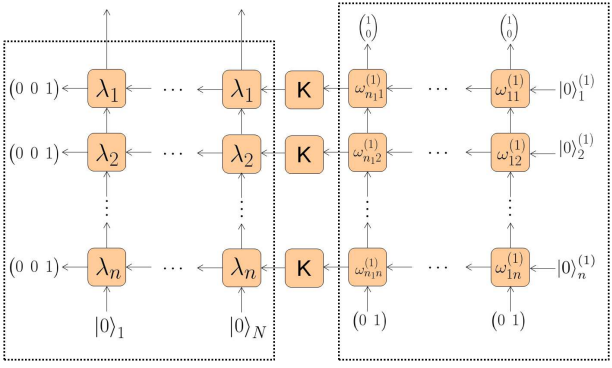}
    \caption{\label{fig:tensor_network}Tensor Network representation}
\end{figure}

Now, we can construct the general tensor network form of the algebraic Bethe ansatz for both representations, as shown in Fig.~\ref{fig:tensor_network},
where we define:
\eq{
&\omega_{ab}\nes = \lambda_a\nes - \lambda_b \\
&\{n, n_1\} = \left\{ \begin{array}{l l}
\{N_e, N_\downarrow\} , &\textrm{Lai representation} \\
\{N_h + N_\downarrow, N_h\} , &\textrm{Sutherland representation}
\end{array}
\right.
}

The tensor network is split into two main parts: the first level Bethe ansatz and the nested Bethe ansatz. The first level and the nested level are connected by contracting the indices $a_1, \cdots , a_n$ of $C_{a_i}$ of the creation operators in the first level with the wavefunction of the nested level, as shown in \eqref{ansatz_state_lai}. The matrix $K$ in Fig.~\ref{fig:tensor_network} (as defined in \eqref{connector_k}) selects the two first level creation operators $\{C_1,C_2\}$ and connects them to the corresponding index of the wavefunction in the nested Bethe ansatz.

The bond dimension of each bond in the tensor network for the first level Bethe ansatz is $3$, while that for the nested level is $2$. Due to the fact that $C_{a_i}$ are creation operators, ...

\section{Grading in terms of tensor networks} \label{sec:grading}

In this section, we explicitly consider the grading for both representations in detail. The tensor product is graded by assigning Grassmann parities to the basis vectors, which represents the fermionic nature of the t-J model. This introduces minus signs which are shown explicitly in \eqref{monodromy_matrix_component_lai} and \eqref{monodromy_matrix_nested_component_suth}. These minus signs are non-local at first glance, as the exponent of the minus sign of each element in the monodromy matrix depends on the parities of the indices to its right. However, in order to perform the approximate contraction of the tensor network (described in Sec.~\ref{sec:approx_contract}) in a sequential manner, we have to localize these minus signs. We have devised two ways to do this as described in the following.

\subsection{Method 1} \label{sec:grading1}
In this method we shall write the monodromy matrices in a recursive form such that the minus signs are included locally in the $L$ operators. Using such a representation in the form of matrices allows us to contract the tensor network efficiently, especially in languages like Matlab which matrix computations are designed for speed.

\subsubsection{Lai representation}
In Lai representation, the graded tensor products in the first level Bethe ansatz produce non-local minus signs as shown in \eqref{monodromy_matrix_component_lai}. However, since the nested Bethe ansatz consist of a system of two fermions (in which the minus signs cancel), the graded tensor products do not produce any explicit (non-local) minus signs.

We introduce the following notation:
\eq{
&\vep_{k} = \ep_{\alpha_{k}} + \ep_{\beta_{k}} \\
&\left. L_{k}(\lambda)_{\alpha_{k}\beta_{k}}^{ab} \right\vert_{\vep_{k}=y} = 
L_{k}(\lambda)_{\alpha_{k}\beta_{k}}^{ab} \delta_{\vep_{k},y} , \qquad y = 0,1
}
The delta function picks out only the quantum operators of the desired Grassmann parity ($\vep_{k} = 0\textrm{ or }1$). In Lai representation, the fermionic ($\vep_k = 1$) operators are $C_a$ and $B_b$ in \eqref{monodromy_matrix_abcd} ($a,b=1,2$), and the rest are bosonic ($\vep_k = 0$). The original $L$ operator is simply expressed by $L_k(\lambda) = \left. L_{k}(\lambda)\right\vert_{\vep_{k}=0} + \left. L_{k}(\lambda)\right\vert_{\vep_{k}=1}$.
We define the following primed $L$ operator and monodromy matrix:
\begin{widetext}
\eq{
&L^\prime_k(\lambda)_{\alpha\beta}^{ab} = L_k(\lambda)_{\alpha\beta}^{ab} (-1)^{\ep_\alpha}\\
&\{[T_L^\prime(\lambda)]^{ab} \}_{\substack{\alpha_1 \cdots \alpha_L \\ \beta_1 \cdots \beta_L}} = L_L^\prime(\lambda)^{ac_L}_{\alpha_L \beta_L} L_{L-1}^\prime(\lambda)^{{c_{L}}{c_{L-1}}}_{\alpha_{L-1} \beta_{L-1}} \cdots 
L_{1}^\prime(\lambda)^{c_2c_{1}}_{\alpha_{1} \beta_{1}} (-1)^{\sum_{j=2}^L (\ep_{\alpha_j} + \ep_{\beta_j}) \sum_{i=1}^{j-1} \ep_{\alpha_i}}
\label{monodromy_matrix_primed_component_lai}
}
Now, we can write \eqref{monodromy_matrix_component_lai} in a recursive form that allows the minus signs to be localized:
\eq{
\begin{pmatrix}
\{[T_{k+1}(\lambda)]^{ab} \}_{\substack{\alpha_1 \cdots \alpha_{k+1} \\ \beta_1 \cdots \beta_{k+1}}} \\
\{[T_{k+1}^\prime(\lambda)]^{ab} \}_{\substack{\alpha_1 \cdots \alpha_{k+1} \\ \beta_1 \cdots \beta_{k+1}}}\end{pmatrix}
=
\begin{pmatrix}
\left. L_{k+1}(\lambda)_{\alpha_{k+1}\beta_{k+1}}^{ac_{k+1}} \right\vert_{\vep_{k+1}=0} & \left. L_{k+1}(\lambda)_{\alpha_{k+1}\beta_{k+1}}^{ac_{k+1}} \right\vert_{\vep_{k+1}=1} \\
\left. L_{k+1}^\prime(\lambda)_{\alpha_{k+1}\beta_{k+1}}^{ac_{k+1}} \right\vert_{\vep_{k+1}=1} & \left. L_{k+1}^\prime(\lambda)_{\alpha_{k+1}\beta_{k+1}}^{ac_{k+1}} \right\vert_{\vep_{k+1}=0} \\
\end{pmatrix}
\begin{pmatrix}
\{[T_{k}(\lambda)]^{c_{k+1}b} \}_{\substack{\alpha_1 \cdots \alpha_{k} \\ \beta_1 \cdots \beta_{k}}} \\
\{[T_{k}^\prime(\lambda)]^{c_{k+1}b} \}_{\substack{\alpha_1 \cdots \alpha_{k} \\ \beta_1 \cdots \beta_{k}}}\end{pmatrix} \label{monodromy_matrix_recursive_lai}
}

The minus signs are absorbed locally into the definition of $L^\prime_k(\lambda)$. The $L$ operators are now embedded in a larger matrix space, which we call the external matrix space. To use this construction to handle the grading, we would have to alter our tensor network so to include the external matrix space.
\begin{figure}[h]
    \includegraphics[width=0.8\textwidth]{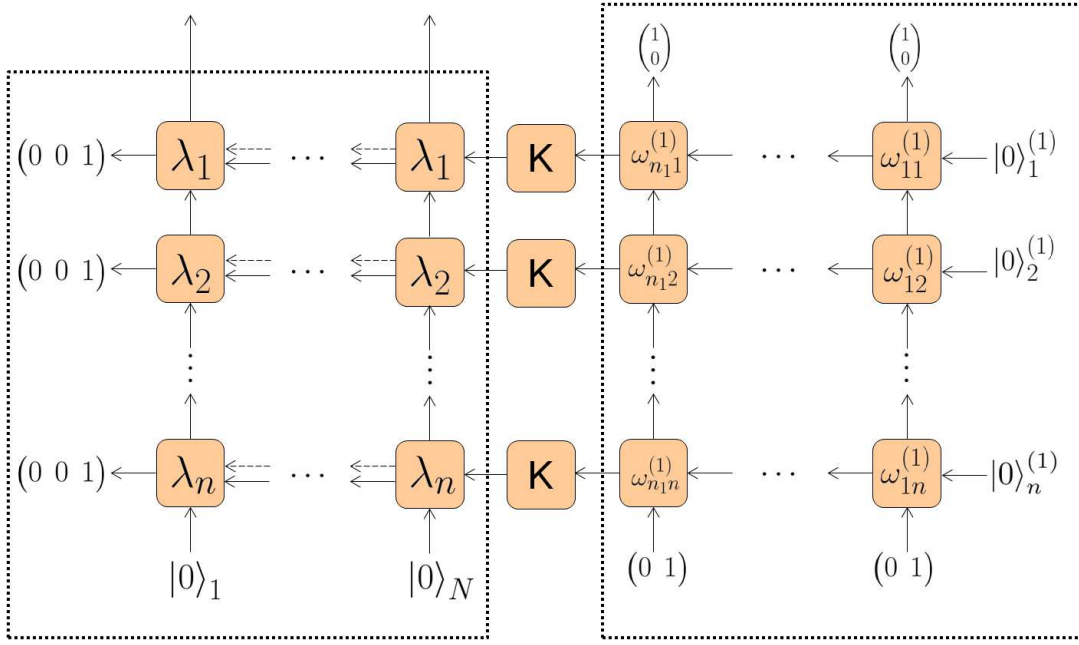}
    \caption{\label{fig:tensor_network_graded_lai}Graded Tensor Network for Lai representation}
\end{figure}
\end{widetext}

$K^\prime$ is defined as:
\eq{
K^\prime = \begin{pmatrix}
1\\1
\end{pmatrix}
\otimes
\begin{pmatrix}
1&0\\
0&1\\
0&0
\end{pmatrix}
}
The boundary vectors on the left of Fig.~\ref{fig:tensor_network_graded_lai} and $K^\prime$ live in the space $V^{(0\vert 2)}\otimes V^{(1\vert 2)}$, where the first space $V^{(0\vert 2)}$ is the external matrix space and the second space $V^{(1\vert 2)}$ is the matrix space.

\subsubsection{Sutherland representation}
For Sutherland representation, the graded tensor products in both the first level and nested Bethe ansatz produce minus signs. The minus signs produced by the tensor product in the first level Bethe ansatz is exactly the same as in Lai representation as shown in \eqref{monodromy_matrix_component_lai}. However, due to the choice of grading in Sutherland representation, the fermionic ($\vep_k = 1$) operators are $B_1$, $C_1$, $A_{12}$ and $A_{21}$ in \eqref{monodromy_matrix_abcd}, and the rest are bosonic ($\vep_k = 0$). Nevertheless, the form of the recursion relation of the first level monodromy matrix for Sutherland representation is exactly the same as \eqref{monodromy_matrix_recursive_lai} in Lai representation.

Now, for the graded tensor product \eqref{monodromy_matrix_nested_component_suth} in the nested Bethe ansatz, we introduce the following:
\begin{widetext}
\eq{
&L^{(1)\prime}_k(\lambda)_{\alpha\beta}^{ab} = L_k^{(1)}(\lambda)_{\alpha\beta}^{ab} (-1)^{\ep_\alpha}\\
&\{[T_L^{(1)\prime}(\lambda)]^{ab} \}_{\substack{\alpha_1 \cdots \alpha_L \\ \beta_1 \cdots \beta_L}} = L_L^{(1)\prime}(\lambda)^{ac_L}_{\alpha_L \beta_L} L_{L-1}^{(1)\prime}(\lambda)^{{c_{L}}{c_{L-1}}}_{\alpha_{L-1} \beta_{L-1}} \cdots 
L_{1}^{(1)\prime}(\lambda)^{c_2c_{1}}_{\alpha_{1} \beta_{1}} (-1)^{\sum_{j=2}^L (\ep_{\alpha_j} + \ep_{\beta_j}) \sum_{i=1}^{j-1} (\ep_{\alpha_i}+1)}
\label{monodromy_matrix_primed_component_suth}
}

Now, we can write \eqref{monodromy_matrix_nested_component_suth} in a recursive form that allows the minus signs to be localized:
\eq{
\begin{pmatrix}
\{[T_{k+1}\nes(\lambda)]^{ab} \}_{\substack{\alpha_1 \cdots \alpha_{k+1} \\ \beta_1 \cdots \beta_{k+1}}} \\
\{[T_{k+1}^{(1)\prime}(\lambda)]^{ab} \}_{\substack{\alpha_1 \cdots \alpha_{k+1} \\ \beta_1 \cdots \beta_{k+1}}}\end{pmatrix}
=
\begin{pmatrix}
\left. L_{k+1}\nes(\lambda)_{\alpha_{k+1}\beta_{k+1}}^{ac_{k+1}} \right\vert_{\vep_{k+1}=0} & \left. L_{k+1}\nes(\lambda)_{\alpha_{k+1}\beta_{k+1}}^{ac_{k+1}} \right\vert_{\vep_{k+1}=1} \\
\left. L_{k+1}^{(1)\prime}(\lambda)_{\alpha_{k+1}\beta_{k+1}}^{ac_{k+1}} \right\vert_{\vep_{k+1}=1} & \left. L_{k+1}^{(1)\prime}(\lambda)_{\alpha_{k+1}\beta_{k+1}}^{ac_{k+1}} \right\vert_{\vep_{k+1}=0} \\
\end{pmatrix}
\begin{pmatrix}
\{[T_{k}\nes(\lambda)]^{c_{k+1}b} \}_{\substack{\alpha_1 \cdots \alpha_{k} \\ \beta_1 \cdots \beta_{k}}} \\
\{[T_{k}^{(1)\prime}(\lambda)]^{c_{k+1}b} \}_{\substack{\alpha_1 \cdots \alpha_{k} \\ \beta_1 \cdots \beta_{k}}}\end{pmatrix}
}

The minus signs in the nested Bethe ansatz are absorbed locally into the definition of $L^{(1)\prime}_k(\lambda)$. To use this construction to handle the grading, we would have to alter our tensor network so to include the external matrix space (in both the first level and nested Bethe ansatz for Sutherland representation).

\begin{figure}[h]
    \includegraphics[width=0.8\textwidth]{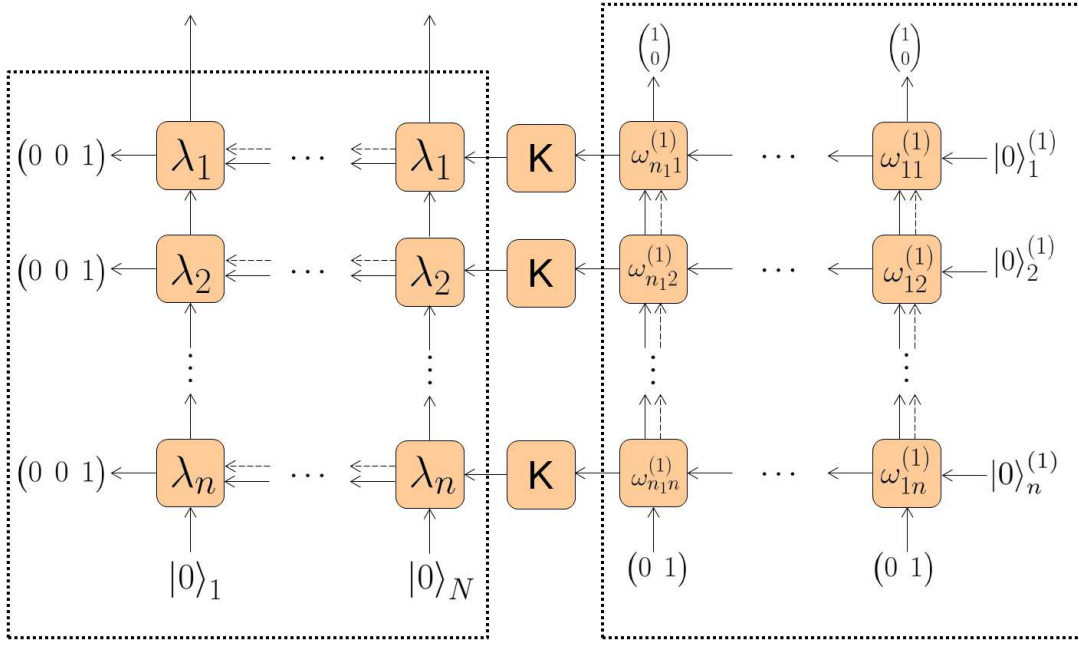}
    \caption{\label{fig:tensor_network_graded_suth}Graded Tensor Network for Sutherland representation}
\end{figure}
\end{widetext}

The boundary vectors on the left of Fig.~\ref{fig:tensor_network_graded_suth} and $K^\prime$ live in the space $V^{(1\vert 1)}\otimes V^{(1\vert 2)}$, where the first space $V^{(1\vert 1)}$ is the external matrix space and the second space $V^{(1\vert 2)}$ is the matrix space, of the first level $L\nes$ operators. The boundary vectors to the top and bottom of the nested Bethe ansatz live similarly in the space $V^{(1\vert 1)}\otimes V^{(1\vert 1)}$, where the first space is the external matrix space and the second space is the matrix space, of the nested $L\nes$ operators.

\subsection{Method 2} \label{sec:grading2}

\subsubsection{Lai representation}

In Lai representation, the grading of the first level Bethe network can also be handled by adding an extra bond that carries the parity information of the indices, denoted by the dotted lines in Fig.~\ref{fig:tensor_network_graded_lai}. The parity bond $p_m$ at the $m^{th}$ site satisfies the relation $p_m = p_{m-1} + \ep_{k_m} \, (\textrm{mod 2})$, where $p_0 = 0$. In addition, these parity bonds, which store local information about the minus signs of~\eqref{monodromy_matrix_component_lai}, satisfy the recursive relation
\eq{
(-1)^{\sum_{j=2}^m (\ep_{k_j} + \ep_{l_j}) \sum_{i=1}^{j-1} \ep_{k_i}}
&= (-1)^{\sum_{j=2}^{m-1} (\ep_{k_j} + \ep_{l_j}) \sum_{i=1}^{j-1} \ep_{k_i}} \nb \\
&\;\;\;\; \times (-1)^{ (\ep_{k_m} + \ep_{l_m}) p_m} \label{parity_recursion}
}

As such, in the tensor network picture with grading, each $L$ operator $L_m$ becomes a tensor with $6$ indices: $2$ horizontal indices of dimension $3$ describing the matrix space, $2$ vertical indices $k_m$ and $l_m$ of dimension $3$ describing the physical space and $2$ parity indices $p_{m-1}$ and $p_m$ of dimension $2$. Because of the recursive relation~\eqref{parity_recursion}, the nonlocal minus signs of~\eqref{monodromy_matrix_component_lai} can be reproduced by multiplying each $L$ operator with $(-1)^{ (\ep_{k_m} + \ep_{l_m}) p_m}$.

\subsubsection{Sutherland representation}
In Sutherland representation, both the first and the nested level Bethe network are graded, and they are handled by adding an extra bond that carries the parity information of the indices, denoted by the dotted lines in both levels of the Bethe ansatz in Fig.~\ref{fig:tensor_network_graded_suth}. As before, the parity bond $p_m$ at the $m^{th}$ site satisfies the relation $p_m = p_{m-1} + \ep_{k_m} \, (\textrm{mod 2})$, where $p_0 = 0$, such that the minus signs of~\eqref{monodromy_matrix_component_lai} can be localized.

\subsection{Equivalence of the two methods}
Upon joining the additional parity bonds (of dimension $2$) in the second method with the bonds in the matrix space (of dimension $3$) of the original tensor network, the $L$ operators are now tensors of $6$ by $6$ in the matrix space and $3$ by $3$ in the physical space, which has the same dimensions as that of the $L$ operators of the first method. These two methods will then give rise to exactly the same tensor network, producing equivalent tensors (up to a unitary transformation). The first method can thus be simply considered as an explicit formulation of the joining of the parity bonds with the original bonds in the matrix space in the second method.

\section{Approximate contraction of the tensor network} \label{sec:approx_contract}

\begin{figure}[h]
    \includegraphics[width=0.45\textwidth]{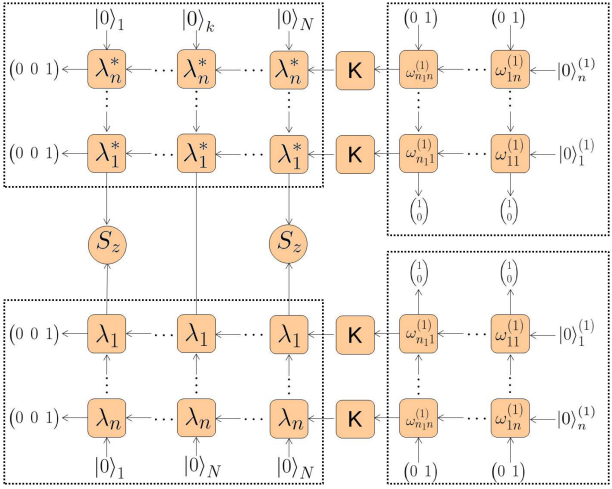}
    \caption{\label{fig:tensor_network_expectation}Tensor Network calculation of expectation values}
\end{figure}

The calculation of expectation values with respect to a Bethe eigenstate of the form of \eqref{ansatz_state_lai} is a considerably complex problem, because it requires the contraction of the tensor network depicted in Fig.~\ref{fig:tensor_network_expectation}. 
A tensor network with such a structure also appears in connection with the calculation of partition functions of two-dimensional classical systems and one-dimensional quantum systems and the calculation of expectation values with respect to PEPSs. The complexity of contracting this network scales exponentially with the number of rows $M$ or columns $N$ (depending on the direction of contraction), which renders exact calculations infeasible.

Following Murg et al.\cite{murg12}, to circumvent this problem, we attempt to perform the contraction in an approximative numerical way: the main idea is to consider the network in Fig.~\ref{fig:tensor_network} as the time evolution of MPOs ($L$ operators) acting on MPSs in a sequential order. 
After each evolution step, the state remains an MPS, but the virtual dimension is increased, by a factor of $3$ (first level) or $2$ (nested level). Thus, we approximate the MPS after each evolution step by a MPS with smaller virtual dimension. Of course, we must exercise caution, as the creation operators are not unitary and the intermediate states of the evolution can be nonphysical (i.e., they might have to be represented by an MPS with high virtual dimension). 
We choose the order of contraction to be such:
\begin{enumerate}
\item In the nested Bethe ansatz, act the $n_1$ nested creation operators $C\nes(\lambda_{n_1}\nes) \cdots C\nes(\lambda_1\nes)$ on the initial MPS $\ket{0}\nes$ sequentially, contracting approximately to get an MPS at each step, to produce a boundary MPS on the right of the first level Bethe ansatz.
\item Now, in the first level Bethe ansatz, $n$ first level creation operators $C(\lambda_{n}) \cdots C(\lambda_1)$ on the initial MPS $\ket{0}$ sequentially, contracting approximately to get an MPS at each step, with the right end of the first level Bethe ansatz terminated by the boundary MPS produced in the first step.
\end{enumerate}

At each step in the above contraction process, we let 
\eq{
&\ket{\Psi_m} = C_{a_m}(\lambda_{m})\ket{\tilde{\Psi}_{m-1}} , \,\, m = 1, \cdots, n\\
&\ket{\Psi_{m_1}}\nes = C\nes(\lambda_{m_1}\nes)\ket{\tilde{\Psi}_{m_1-1}\nes} , \,\, m_1 = 1, \cdots, n_1
}
where
\eq{
\ket{\tilde{\Psi}_0} = \ket{0} , \quad \ket{\tilde{\Psi}_0}\nes = \ket{0}\nes
}

At each step of the first level Bethe ansatz, $\ket{\Psi_m}$ is approximated by the MPS $\ket{\tilde{\Psi}_m}$ that has maximal bond dimension $D$ and is closest to $\ket{\tilde{\Psi}_m}$. In other words, we try solve the minimization problem
\eq{
K := \left\| \ket{\Psi_m} - \ket{\tilde{\Psi}_m} \right\|^2 \rightarrow \textrm{min}
}
by optimizing over all matrices of the MPS $\ket{\tilde{\Psi}_m}$. (This minimization is done in the same manner for the nested Bethe ansatz.) This minimization problem also appears in the context of numerical calculation of expectation values with respect to PEPSs, calculation of partition functions, and (imaginary) time evolution of one-dimensional quantum systems.
In this way, the MPS approximation of the Bethe state is obtained for the whole tensor network. 
The error of the approximation is well controlled in the sense that the expectation value of the energy can always be calculated with respect to the approximated MPS $\ket{\tilde{\Psi}_m}$ and compared to the exact energy available from the Bethe ansatz.

There is a (mathematical) degree of freedom that can be used to improve the approximation. This degree of freedom is due to the fact that the set of $\{\{\lambda\}, \{\lambda\nes\}\}$ encode information about physical quantities and the ordering of the them should not change the final wavefunction produced. That is, permutation of order of applying the creation operators through permutation of the set of $\{\{\lambda\}, \{\lambda\nes\}\}$ will not change the final wavefunction.
However, the intermediate states are a priori not physical ground states; i.e., there is no reason for them to lie in the set of MPS with low bond dimension.
Even so, similar to that which is noted in \cite{murg12}, there is always an ordering of the set of $\lambda$'s such that the intermediate states contain as little entanglement as possible. We then use that ordering for doing the approximation.

\section{Numerical Results}

Using the previously described method, we have obtained numerical results for the t-J model with periodic boundary
conditions. We chose to implement the tensor network representation of the Sutherland representation, since, near half filling, its Bethe ansatz equations are more well behaved numerically, and its tensor network is smaller. We chose to use the first method of implementing grading as the explicit construction of the matrices is more easily checked for errors. As a proof of principle, we obtain the correlation functions of eigenstates on lattices of length $18$ as presented below. Calculations for lattices of larger length can be achieved through consideration of symmetries, or using mathematical packages which can extend the limit of machine precision.

\subsection{Electron correlator}
\begin{figure}[h]
    \includegraphics[width=0.45\textwidth]{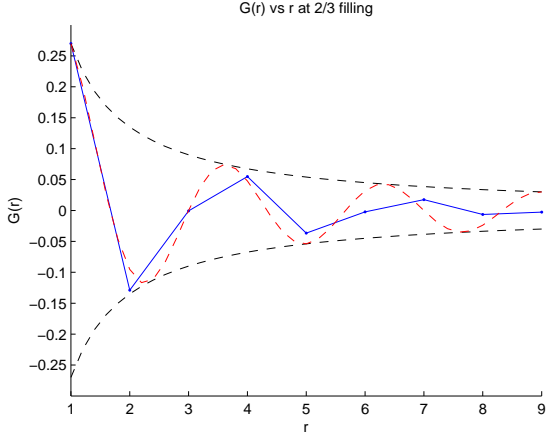}
    \caption{\label{fig:ecorr_up1}Spin-up correlator at 2/3 filling for ground state}
\end{figure}

The asymptotic behavior of the spin correlator is predicted by conformal field theory to be 
\eq{
G_\sigma(r) = \left\langle c_\sigma^\dagger(r) c_\sigma(0) \right\rangle \propto r^{-\eta} cos(k_F r)
}
where $\eta$ and $k_F$ are as defined in \cite{kawakamiyang91}.
This is strongly supported by our results, as can be gathered from Fig.~\ref{fig:ecorr_up1}.

\subsection{Singlet pair superconducting correlators}
\begin{figure}[ht]
    \includegraphics[width=0.45\textwidth]{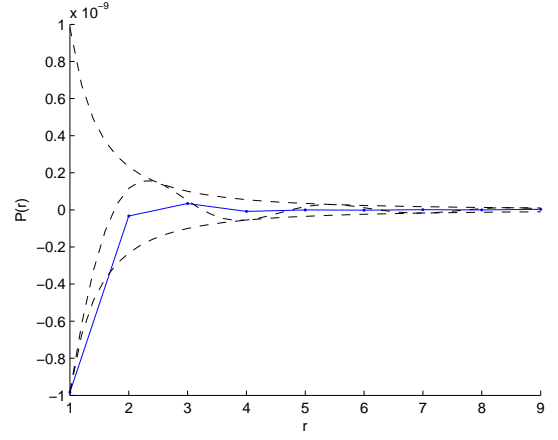}
    \caption{\label{fig:singlet_gs1}Singlet pair superconducting correlator at 2/3 filling for ground state}
\end{figure}
The asymptotic behavior of the singlet pair correlator is predicted by conformal field theory to be 
\eq{
P_s(r) \left\langle c_\uparrow^\dagger(r+1) c_\downarrow^\dagger(r) c_\uparrow(1) c_\downarrow(0) \right\rangle  \propto r^{-\beta_s} cos(2 k_F r)
}
where $\beta_s$ and $k_F$ are as defined in \cite{kawakamiyang91}.
This is strongly supported by our results, as can be gathered from Fig.~\ref{fig:singlet_gs1}.

\subsection{Spin correlator}
\begin{figure}[ht]
    \includegraphics[width=0.45\textwidth]{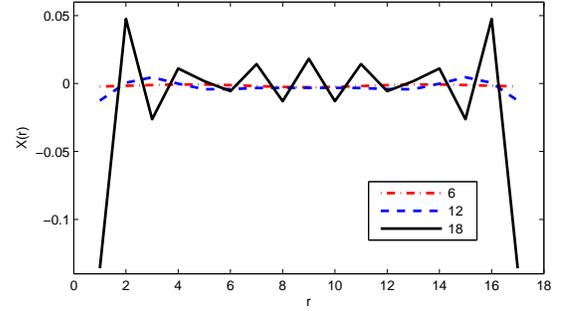}
    \caption{\label{fig:spin_ctrip1}Spin correlator at various filling for charge triplet state - the number in the legend shows $N$, the total number of particles}
\end{figure}
The spin correlator is defined as:
\eq{
\chi(r) = \left\langle S_z(r) S_z(0) \right\rangle \;\; , \;\; S_z(r) = (n_\uparrow(r) - n_\downarrow(r))
}
We consider the charge triplet state and calculate its spin correlator, as shown in Fig.~\ref{fig:spin_ctrip1}. It shows that an interesting trend that as we increase the filling towards half filling, the variation of the spin correlator increases, and that it tends toward a zigzag pattern that alternates between the even and odd lattice sites at half filling.

\subsection{Charge density correlator}
\begin{figure}[ht]
    \includegraphics[width=0.5\textwidth]{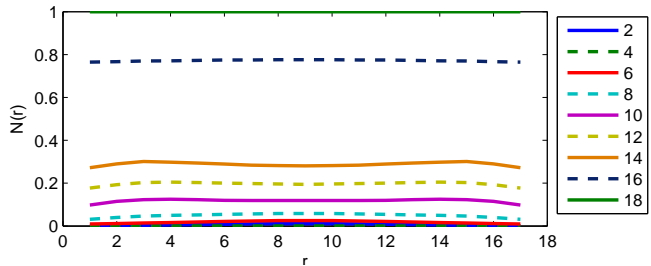}
    \caption{\label{fig:num_ctrip1}Charge density correlator at various filling for charge triplet state (color online) - the number in the legend shows N, the total number of particles}
\end{figure}
\begin{figure}[ht]
    \includegraphics[width=0.5\textwidth]{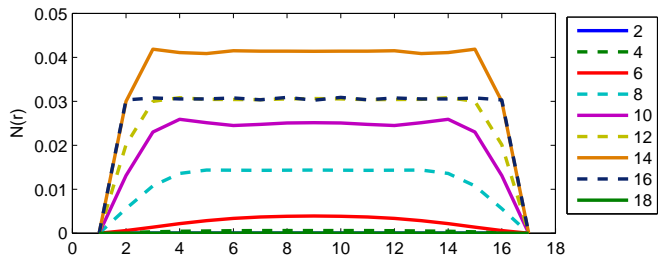}
    \caption{\label{fig:num_ctrip0}Normalized charge density correlator at various filling for charge triplet state (color online) - the number in the legend shows $N$, the total number of particles}
\end{figure}

The charge density correlator is defined as:
\eq{
N(r) = \left\langle n(r) n(0) \right\rangle \;\; , \;\; n(r) = (n_\uparrow(r) + n_\downarrow(r))
}
We consider the charge triplet state and calculate its charge density correlator, as shown in Fig.~\ref{fig:num_ctrip1}. It does not fully show the trend of variation across the lattice sites, as it is dominated by the constant term in the correlator. As such, we attempt to ``normalize'' the correlator by setting the correlator of the first site to be zero (by translation), as shown in Fig.~\ref{fig:num_ctrip0}. This clearly shows a trend that as $N$ (total number of particles) increases, the magnitude of the variation of the correlator across the lattice sites increases, until $N=14$ which is reaches a peak, then decreases.

\section{Conclusions}
Summing up, we have presented a method for approximative calculation of expectation values with respect to
Bethe eigenstates of the t-J model. To achieve this, we make use of the fact that a Bethe eigenstate is a product of MPOs applied to an MPS. We systematically reduce the virtual dimension after each multiplication and obtain an MPS with small virtual dimension that can be used for the calculation of any expectation value. As a proof of principle, we have obtained the correlation functions of eigenstates on finite length lattices with our method.

\appendix

\acknowledgments{
V.~M.\ and F.~V.\ acknowledge support from the SFB projects
FoQuS and ViCoM, the European projects Quevadis, and the ERC grant
Querg. V.~E.~K.\ and Y.~Q.~C.\ acknowledge support from the NSF grant DMS-1205422.
}

\end{document}